\documentclass[12pt,preprint]{aastex}

\def\lsim{\lower.5ex\hbox{$\; \buildrel < \over \sim \;$}}
\def\gsim{\lower.5ex\hbox{$\; \buildrel > \over \sim \;$}}

\shorttitle{Evidence for cosmic-ray acceleration in RS Oph}
\shortauthors{Tatischeff \& Hernanz}

\begin{document}



\title{Evidence for nonlinear diffusive shock acceleration of cosmic-rays in the 2006 outburst of the recurrent nova RS Ophiuchi}

\author{V. Tatischeff\altaffilmark{1} and M. Hernanz}
\affil{Institut de Ci\`encies de l'Espai (CSIC-IEEC), Campus UAB, Fac. Ci\`encies, 08193 Bellaterra, Barcelona, Spain}
\email{tatische@csnsm.in2p3.fr, hernanz@ieec.uab.es}

\altaffiltext{1}{Permanent adress: CSNSM, IN2P3-CNRS and Univ 
Paris-Sud, F-91405 Orsay Cedex, France}


\begin{abstract}

Spectroscopic observations of the 2006 outburst of the recurrent nova RS Ophiuchi at both infrared (IR) and X-ray wavelengths have shown that the blast wave has decelerated at a higher rate than predicted by the standard 
test-particle adiabatic shock-wave model. Here we show that the observed evolution of the nova remnant can be explained by the diffusive shock acceleration of particles at the blast wave and the subsequent escape of the highest energy ions from the shock region. Nonlinear particle acceleration can also account for the difference of shock velocities deduced from the IR and X-ray data. The maximum energy that accelerated electrons and protons can have achieved in few days after outburst is found to be as high as a few TeV. Using the semi-analytic model of nonlinear diffusive shock acceleration developed by Berezhko \& Ellison, we show that the postshock temperature of the shocked gas measured with {\it RXTE} PCA and {\it Swift} XRT imply a relatively moderate acceleration efficiency characterized by a proton injection rate $\eta_{\rm inj}$$\gsim$10$^{-4}$. 

\end{abstract}




\section{Introduction}

RS Ophiuchi is a symbiotic recurrent nova with various recorded eruptions in the last century (the last one in 1985), which erupted again recently, on 2006 February 12 (Hirosawa 2006). RS Oph's binary system consists of a white dwarf with mass near the Chandrasekhar limit, and a red giant (RG) companion star. High white dwarf mass and large accretion rate lead to a much shorter recurrence period of outbursts than in classical novae (where the donor is a main sequence star). In addition, the presence of the RG wind in the RS Oph system leads to the generation of an X-ray emitting blast wave that runs into a relatively dense circumstellar medium.

The latest outburst of RS Oph has been observed at various wavelengths, e.g. in radio (O'Brien et al. 2006), IR (Monnier et al. 2006; Das et al. 2006; Evans et al. 2007) and X-rays (Sokoloski et al. 2006; Bode et al. 2006). The X-ray data have allowed to clearly identify the forward shock wave expanding into the RG wind and to estimate the time evolution of its velocity, $v_s$, through the well-known relation for a test-particle strong shock:
\begin{equation}
v_s = \bigg({16 \over 3} {k T_s \over \mu m_H}\bigg)^{0.5}~,
\end{equation}
where $k$ is the Boltzmann constant, $T_s$ is the measured postshock temperature and $\mu m_H$ is the mean particle mass. The X-ray emission has revealed that after an ejecta-dominated, free expansion stage (phase I) lasting $\sim$6~days, the remnant rapidly evolved to display behavior characteristic of a shock experiencing significant radiative cooling (phase III; see Sokoloski et al. 2006; Bode et al. 2006). At day 6 after outburst, however, the shocked material was so hot, $T_s$$\sim$10$^{8}$~K, that its cooling by radiative losses was probably not important for the dynamics of the shock. Thus, the lack or the very short duration of an adiabatic, Sedov-Taylor phase (phase II) differs from the remnant evolution model developed by Bode \& Kahn (1985), O'Brien \& Kahn (1987), and O'Brien et al. (1992) after the 1985 outburst of RS Oph. 

The time-dependence of shock velocity has also been measured by IR spectroscopy, using the observed narrowing of strong coronal emission lines (Das et al. 2006; Evans et al. 2007). Although the general behavior of the shock evolution was found to be consistent with that deduced from the X-ray emission, the shock velocities determined from the IR data are significantly greater than those obtained using eq.~(1) together with the X-ray measurements of $T_s$ (see  Fig.~1). 

In this Letter, we show that production of nonthermal particles at the forward shock through the first-order Fermi acceleration process can be deduced from these observational data. Several observations in the solar system and beyond show that diffusive acceleration of particles can be efficient in collisionless shocks and the backpressure from the energetic ions can strongly modify the shock structure (e.g. Jones \& Ellison 1991). In particular, eq.~(1) is known to underestimate shock velocities when particle acceleration is efficient, because the postshock temperature can be much lower than the test-particle value (Decourchelle et al. 2000; Ellison et al. 2007). We start in \S~2 with a simple description of the dynamical evolution of the blast wave based on IR and X-ray observations, we then calculate in \S~3 the maximum possible energy of particles accelerated in the nova remnant and determine in \S~4 the properties of the cosmic-ray modified shock including the energy carried off by particles escaping the shock region. Our conclusions follow in \S~5.

\section{Shock wave evolution}

The FWZI (full width at zero intensity) of IR emission lines of coronal origin
should provide a good measurement of the shock velocity (Das et al. 2006; Evans et al. 2007). Radio imaging (O'Brien et al. 2006) and IR interferometric observations (Monnier et al. 2006) of the RS Oph remnant have shown departures from spherical symmetry. But given the intermediate angle between the symmetry axis of the observed bipolar structure and the line of sight, $\theta$$\sim$50--60$^\circ$ (O'Brien et al. 2006), the largest blueshifted and redshifted velocities measured in the FWZI should be close to the mean expansion speed of the blast wave. The IR data can be modeled at first approximation by (dashed line in Fig.~1) $v_s(t) = 4300 \tau^{\alpha_v} {\rm~km~s^{-1}}$, where $t$ is the time after outburst, $\tau$=$t/t_1$ with $t_1$=6~days, and $\alpha_v$=0 (-0.5) for $t$$\leq$$t_1$ ($t$$>$$t_1$). The decay of $v_s$ as $t^{-0.5}$ for $t$$>$$t_1$ can be expected from a well-cooled shock. We see that the velocities deduced from the X-ray data are consistent with such a decay (dotted line in Fig.~1), although they are $\sim$1.7 times lower than the velocities determined from the IR lines. 

The radius of the shock front, which for simplicity we assume to be spherical, is easily obtained by integration of $v_s(t)$: $r_s(t) = 2.23\times10^{14} [(1-2\alpha_v) \tau^{\alpha_v+1} + 2\alpha_v] {\rm~~cm}$. We do not consider the earliest phase of the outburst, when the shock wave traversed the binary system. Given the binary separation of $\sim$1.5~AU (Fekel et al. 2000), the free expansion of the ejecta into the unperturbed RG wind started at $t$$\sim$$t_0$=1~day. The outer radius of the RG wind is $r_{\rm out}$=$u_{\rm RG}$$\Delta t$, with $u_{\rm RG}$$\cong$10--20~km~s$^{-1}$ the terminal speed of the RG wind and $\Delta t$=21.04~yr the elapsed time between the 1985 and 2006 outbursts. The outer radius was reached by the forward shock at $t_2$$\cong$24--72~days after the 2006 outburst. 

The density of the RG wind as a function of radius is given by $\rho_W(r) = \dot{M}_{\rm RG} /(4 \pi r^2 u_{\rm RG})$, where $\dot{M}_{\rm RG}$ is the RG mass-loss rate. It can be estimated from the X-ray photoelectric absorption  measured with {\it Swift} XRT (Bode et al. 2006) . We use equation~(4) of Bode et al. (2006) to fit the measured absorbing column density. We do not take into account in this fit the data taken at $t$=3.17~days, because the early X-ray emission could partially originate from the reverse shock running into the ejecta. We obtain $\dot{M}_{\rm RG}$/$u_{\rm RG}$$\approx$4$\times$10$^{13}$~g~cm$^{-1}$, which is $\sim$5 times higher than the value estimated by O'Brien et al. (1992) for the 1985 outburst. Part of the difference is due to the larger ejecta velocity considered here. 

Relatively large magnetic field of stellar origin is expected to pre-exist in the RG wind. Assuming that turbulent motions in the wind amplify the magnetic field $B_W$ up to the equipartition value (Bode \& Kahn 1985), we have $B_W = (8 \pi \rho_W kT_W / \mu m_H)^{0.5}$, where $T_W$ is the wind temperature, which we assume to be uniform throughout the wind and equal to 10$^4$~K. In the vicinity of the shock front, further magnetic field amplification is expected to occur due to various interactions between accelerated particles and plasma waves (e.g. Lucek \& Bell 2000). Assuming a time-independent amplification factor $\alpha_B$ (expected to be of the order of a few), the magnetic field just ahead of the shock, $B_0$=$\alpha_B$$B_W$, can be evaluated using the above relations for the RG wind density and shock radius: 
\begin{equation}
B_0(t) = 0.047 \alpha_B [(1-2\alpha_v) \tau^{\alpha_v+1} + 2\alpha_v ]^{-1}  {\rm~~G}.
\end{equation}
Given these estimates of $v_s$, $r_s$, $\rho_0$=$\rho_W(r_s)$, and $B_0$, we can now study the effects of particle acceleration at the blast wave. 

\section{Acceleration rate and maximum particle energy}

The rate of momentum gain of nonthermal particles diffusing in the vicinity of a shock wave is given by (e.g. Lagage \& Cesarsky 1983)
\begin{equation}
\bigg({dp \over dt}\bigg)_{\rm acc}= {p(u_0 - u_2) \over 3} 
\bigg({\kappa_0(p) \over u_0}+{\kappa_2(p) \over u_2}\bigg)^{-1}~,
\end{equation}
where $u_0$ ($u_2$) is the upstream (downstream) component of flow speed normal to the shock in its rest frame and $\kappa_0$ ($\kappa_2$) is the upstream (downstream)  spatial diffusion coefficient in the direction normal to the shock. Here and elsewhere, the subscript "0" ("2") implies quantities determined far upstream\footnote{By far upstream, we mean ahead of the upstream shock precursor induced by the backpressure of energetic particles.} (downstream) from the shock front, as in Berezhko \& Ellison (1999). To  estimate the spatial diffusion coefficient, $\kappa$=$\lambda v/3$, we assume the scattering mean free path $\lambda$ of all particles of speed $v$ to be 
 $\lambda =  \eta_{\rm mfp} r_g$ (Baring et al. 1999; Ellison et al. 2000), where $\eta_{\rm mfp}$ is a constant and $r_g$=$pc/(QeB)$ is the particle gyroradius, $c$ being the speed of light, $Q$ the charge number, and $-e$ the electronic charge. The minimum value $\eta_{\rm mfp}$=1 corresponds to the B\"ohm limit.

Because the wind magnetic field should be largely azimuthal, the acceleration efficiency could be significantly reduced as compared with that in a quasi-parallel shock (shock normal parallel to the field direction). A full treatment of this question is beyond the scope of this Letter (see, e.g., Kirk et al. 1996). For simplicity, we consider the shock to be quasi-parallel and allow for large values of $\eta_{\rm mfp}$ to account for the possible reduction of the acceleration efficiency.

Given the expected field compression $B_2$=$r_{\rm tot}$$B_0$, where $r_{\rm tot}$=$u_0$/$u_2$ is the total compression ratio of the shock, we have $\kappa_2(p)$=$\kappa_0(p)$/$r_{\rm tot}$. The rate of energy gain for ultrarelativistic particles ($v$=$c$) can then be written from eq.~(3) as 
\begin{equation}
\bigg({dE \over dt}\bigg)_{\rm acc} = 50.0{r_{\rm tot}-1 \over r_{\rm tot}} {Q \over \eta_{\rm mfp}} B_0  \bigg({v_s \over 10^3 {\rm~km~s^{-1}}}\bigg)^2 {\rm~MeV~s^{-1}},
\end{equation}
where $B_0$ is in units of Gauss. Acceleration can only occur for particles whose acceleration rate is lower than their energy loss rate. Concentrating here on nucleons, we can safely neglect the collision energy losses. Adiabatic losses, however, can limit the maximum energy $E_{\rm max}$ that particles can acquire during the shock lifetime. The rate of momentum loss due to adiabatic deceleration of the nonthermal particles in the expanding flow can be written as 
$(dp/dt)_{\rm ad}= p v_s (r_{\rm tot}-1) / (3 r_s r_{\rm tot})$ 
(e.g. V\"olk \& Biermann 1988),
which gives for the energy loss rate of ultrarelativistic particles accelerated in the nova remnant
\begin{equation}
\bigg({dE \over dt}\bigg)_{\rm ad}=0.64 {r_{\rm tot}-1 \over r_{\rm tot}} E_{\rm TeV} [(1-2\alpha_v) \tau + 2\alpha_v\tau^{-\alpha_v}]^{-1} {\rm~MeV~s^{-1}},
\end{equation}
where $E_{\rm TeV}$ is the particle kinetic energy in TeV. By equalling the adiabatic loss rate with the acceleration rate (eq.~4), one obtains an upper limit on the nonthermal particle energy:
\begin{equation}
E \leq E_{\rm max}=68 {Q \over \eta_{\rm mfp}} \alpha_B \tau^{\alpha_v}~{\rm~~TeV}.
\end{equation}
 
It is likely, however, that high-energy particles can escape the acceleration process before reaching the maximum energy given above. Following Baring et al. (1999), we assume the existence of an upstream free escape boundary (FEB) ahead of the shock, located at some constant fraction $f_{\rm esc}$ of the shock radius: $d_{\rm FEB}=f_{\rm esc}r_s$. The maximum energy that particles can acquire before reaching the FEB is obtained by equalling $d_{\rm FEB}$ to the upstream diffusion length $l_0 \sim \kappa_0 / v_s$. Using the parameters derived in \S~2, one finds for ultrarelativistic particles:
\begin{equation}
E_{\rm max}^{\rm size}=136 f_{\rm esc} {Q \over \eta_{\rm mfp}} \alpha_B \tau^{\alpha_v}~{\rm~~TeV}.
\end{equation}
We see that for $f_{\rm esc}$$<$0.5 this size limitation of the acceleration region gives a more restrictive constraint on the maximum particle energy than that obtained from the adiabatic losses (eq.~6). 

Calculated maximum proton energies are shown in Fig.~2 for $f_{\rm esc}$=0.25 (Baring et al. 1999) and $\alpha_B / \eta_{\rm mfp}$=0.1. The quantity $E_{\rm max}^{\rm age}$ is the maximum proton energy caused by the finite age of the shock, and has been obtained by time integration of $(dE / dt)_{\rm acc}$ from $t_0$ to $t$, adopting $r_{\rm tot}=6.5$ (this value of $r_{\rm tot}$ is close to what we obtain for $t$$>$$t_1$, see Fig.~3c). We have slightly underestimated $E_{\rm max}^{\rm age}$ for the initial nonrelativistic phase by assuming $v$=$c$, but the error is negligible given the very short duration of this phase. The production of proton energies above 1~TeV is a consequence of the high values of $B_0$ (eq.~2) implied by the assumption of equipartition. In Fig.~2, the time for which $E_{\rm max}^{\rm age}$=$E_{\rm max}^{\rm size}$, i.e. the beginning of particle escape from the shock region is $t_{\rm esc}$=5.95~days. This time only depends on $f_{\rm esc}$, but not on $\alpha_B / \eta_{\rm mfp}$ which is a scale factor for both $E_{\rm max}^{\rm age}$ and $E_{\rm max}^{\rm size}$. It is remarkable that for $f_{\rm esc}$=0.25, $t_{\rm esc}$ is very close to the observed transition time $t_1$, which could explain the apparent lack of an adiabatic phase in the remnant evolution. 

\section{Properties of the cosmic-ray modified shock}

Berezhko \& Ellison (1999) have developed a relatively simple model of nonlinear diffusive shock acceleration, which allows to quantify the modification of the shock structure induced by the backreaction of energetic ions.  Although the model strictly applies to plane-parallel, steady state shocks, it has been sucessfully used by Ellison et al. (2000) for evolving supernova remnants. Given the upstream sonic and Alfv\'en Mach numbers of the shock, which can be readily calculated from the parameters derived in \S~2, and the maximum particle energies evaluated in \S~3, both the thermodynamic properties of the shocked gas and the energy spectrum of the accelerated protons (other particle species can be neglected for evaluating the shock modification) are determined by an arbitrary injection parameter $\eta_{\rm inj}$, which is the fraction of total shocked protons in protons with momentum $p$$\geq$$p_{\rm inj}$ injected from the thermal pool into the diffusive shock acceleration process. We used the work of Blasi et al. (2005) to accurately relate the injection momentum $p_{\rm inj}$ to $\eta_{\rm inj}$. 

Calculated temperatures of the postshock gas are shown in Fig.~3a. We see that the temperatures measured with {\it RXTE}/PCA and {\it Swift}/XRT can be well reproduced with $\eta_{\rm inj}$=1.4$\times$10$^{-4}$ and Alfv\'en wave heating of the shock precursor. The latter process is thought to be an important ingredient of cosmic-ray acceleration (McKenzie \& V\"olk 1982) and appears to be required in this case as well to limit the shock compression ratio and acceleration efficiency. For $\eta_{\rm inj}$=10$^{-5}$, the test-particle approximation applies and the standard relation between $v_s$ and $T_s$ (eq.~1) overestimates the temperature. 

The solution shown in Fig.~3a is not unique. For example, an equally good description of the $T_s$ measurements can be obtained with $\eta_{\rm mfp}$=100 and $\eta_{\rm inj}$=1.9$\times$10$^{-4}$. However, all the solutions providing good fits to the data give about the same compression ratio and acceleration efficiency. The latter is shown in Fig.~3b for the same input parameters as in panel~(a). The two quantities plotted in this figure are ({\it thin curves}) the fraction of total energy flux, $F_0$, going into nonthermal particles, and ({\it thick curves}) the fraction $\epsilon_{\rm esc}$ of $F_0$ escaping the shock system via diffusion of the highest energy particles across the FEB\footnote{Although $\epsilon_{\rm esc}$ has been set to zero for $t < t_{\rm esc}$, it is likely than some high-energy particles escape upstream from the shock at all times, because of the low scattering strength of high-momentum particles in self-generated turbulence (see Vladimirov et al. 2006).}. Here, $F_0 = 0.5\rho_0 v_s^3 +[\gamma_g / (\gamma_g-1)] P_0 v_s$, where $\gamma_g$=5/3 is the specific heat ratio of the upstream thermal gas and $P_0$ is the far upstream pressure. We see that for 
$\eta_{\rm inj}$=1.4$\times$10$^{-4}$ and Alfv\'en wave heating, $\epsilon_{\rm esc}$=10--20\%, which means that accelerated particle escape is important for the dynamics of the shock. Compression ratios and postshock pressures calculated for this case are shown in Fig.~3c.

The energy loss rate due to the escape of the highest energy particles can be estimated to be
\begin{equation}
\bigg({dE \over dt}\bigg)_{\rm esc} = 4 \pi r_s^2 \epsilon_{\rm esc} F_0
\approx 2.4\times 10^{38} \bigg({\epsilon_{\rm esc} \over 0.15}\bigg) \tau^{-1.5}~{\rm~~erg~s^{-1}},
\end{equation}
for $t \ge t_{\rm esc} \cong t_1$, where we have neglected the enthalpy term $[\gamma_g / (\gamma_g-1)] P_0 v_s$ in the expression for $F_0$.
For $t \cong t_1$, $(dE/dt)_{\rm esc}$ is $\sim$200 times higher (for the assumed distance of 1.6~kpc; Hjellming et al. 1986) that the bolometric luminosity of the postshock hot plasma (Bode et al. 2006; Sokolovsky et al. 2006), which shows that energy loss via accelerated particle escape is much more effective to cool the shock than radiative losses.

\section{Conclusions}

We have shown that production of nonthermal particles by diffusive acceleration at the blast wave generated in the 2006 outburst of RS Oph can reconcile shock velocities deduced from X-ray data with velocities measured in  broad IR lines of coronal origin, and account for the observed cooling of the shock starting as early as $\sim$6~days after outburst. 

Using a semi-analytic model of nonlinear diffusive shock acceleration, we have constrained the proton injection rate from the measured postshock temperature to be $\eta_{\rm inj}$$\gsim$10$^{-4}$. We believe that the existing high-quality multiwavelength observations of this nova outburst could allow to further test and improve the diffusive acceleration theory. 

To our knowledge, the acceleration of particles to TeV energies in a recurrent nova remnant was not predicted. In a forthcoming paper, we will calculate the high-energy emission generated via interactions of this nonthermal population with the ambient medium. 

\acknowledgments{This work has been partially supported by the grants AGAUR 2006-PIV-10044, MEC AYA2004-06290-C02-01, and NSF PHY05-51164.}




\clearpage

\begin{figure}
\begin{center}
\plotone{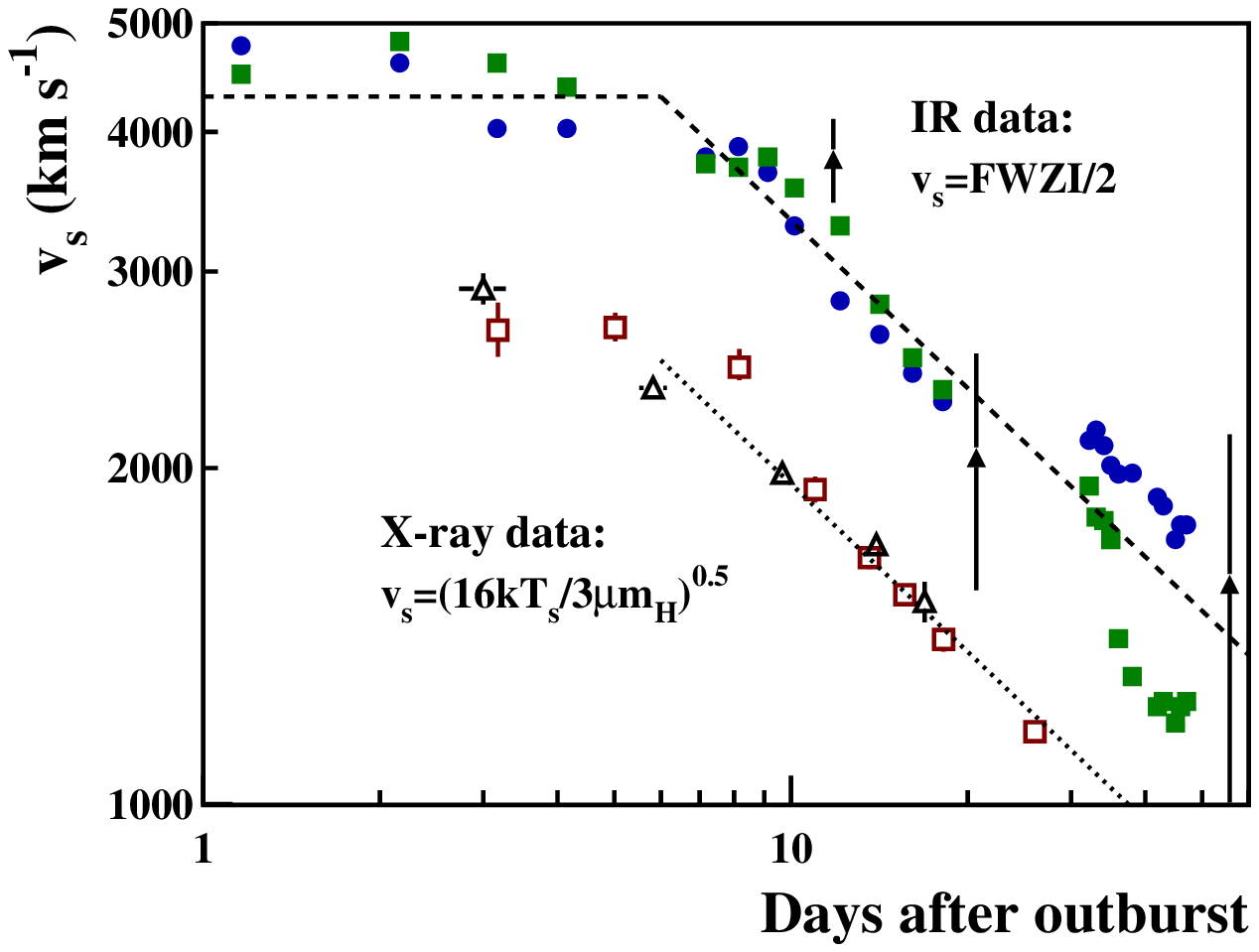}
\figcaption{Time-dependence of forward shock velocity as deduced from FWZI of IR emission lines (filled symbols: squares and circles (Das et al. 2006),  triangles (Evans et al. 2007)) and X-ray measurements of the postshock temperature (open symbols: triangles (Sokoloski et al. 2006), squares (Bode et al. 2006)). The lines are simple fits to the data (see text). \label{fig1}}
\end{center}
\end{figure}

\clearpage

\begin{figure}
\begin{center}
\plotone{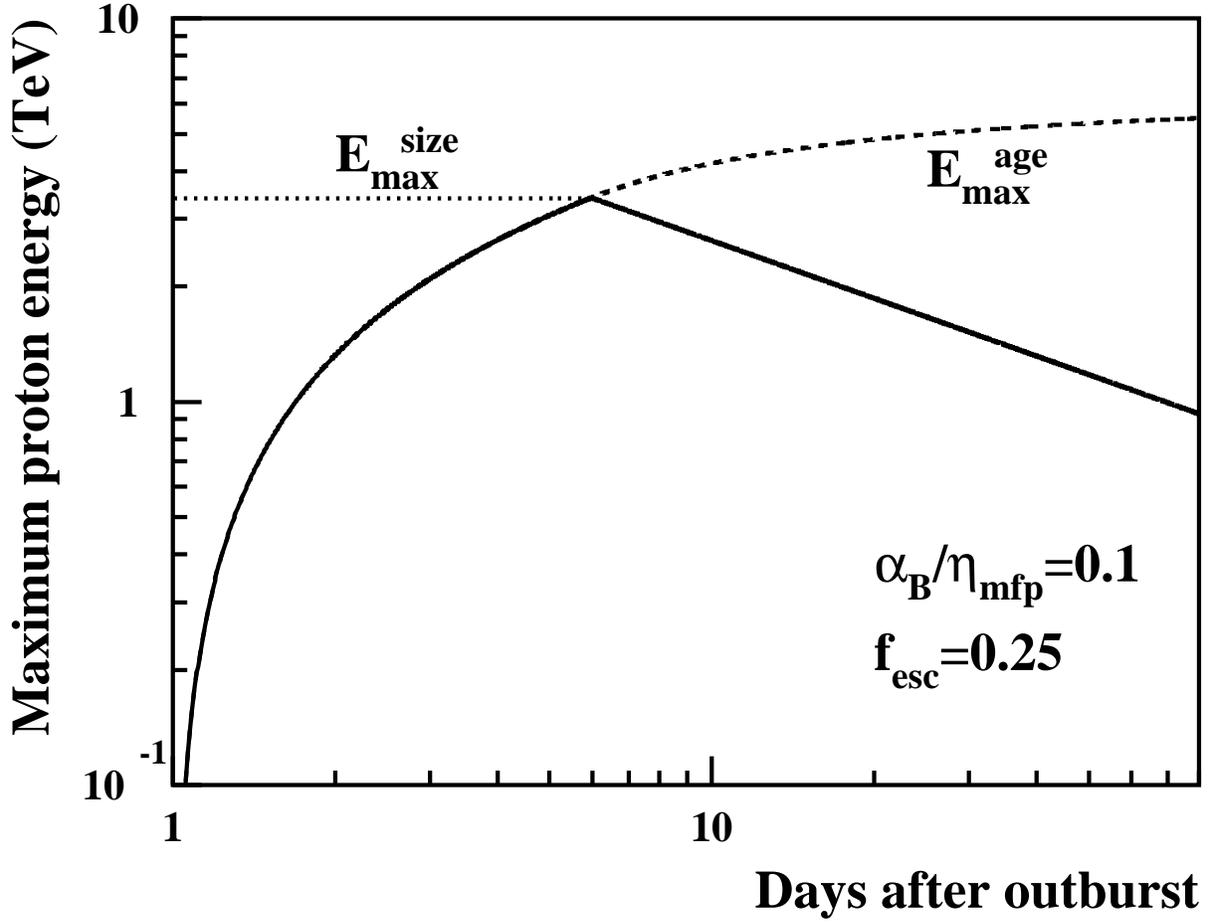}
\figcaption{Estimated maximum proton energy as a function of time after outburst. $E_{\rm max}^{\rm age}$ and $E_{\rm max}^{\rm size}$ are the maximum energies caused by the finite shock age and size, respectively (see text). The solid curve shows the minimum of these two quantities. \label{fig2}}
\end{center}
\end{figure}

\clearpage

\begin{figure}
\begin{center}
\includegraphics[width=10.5cm]{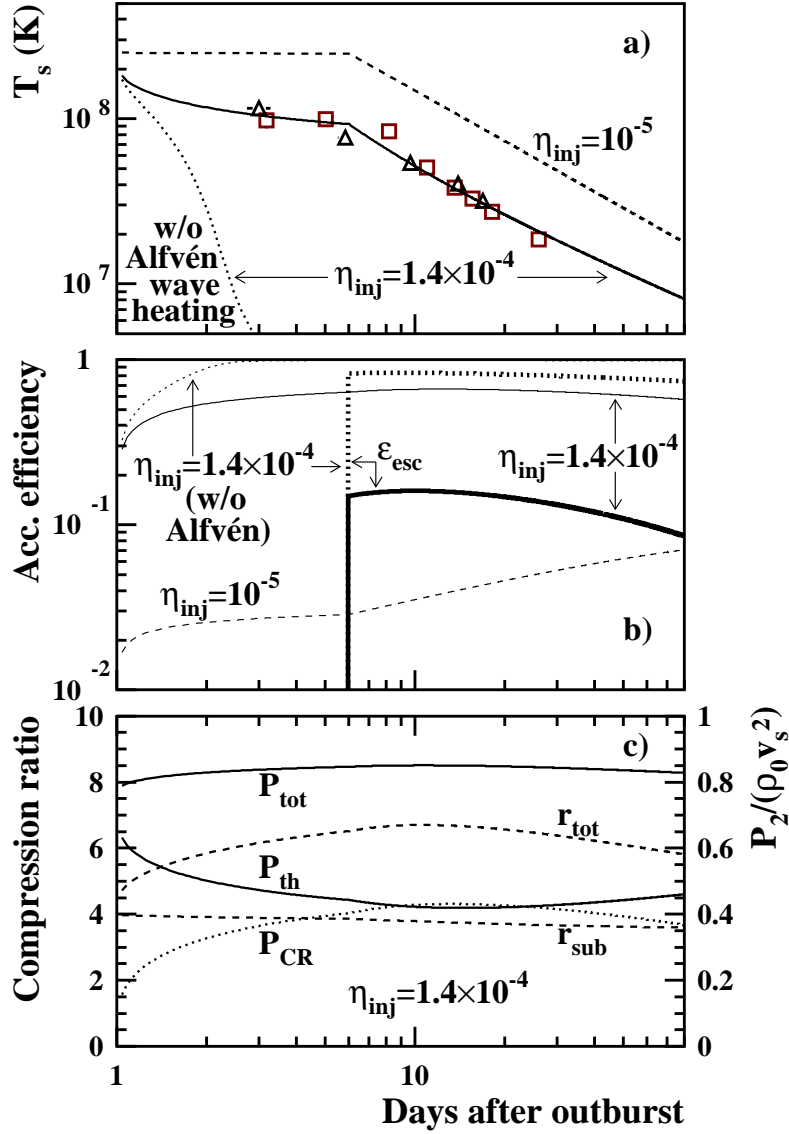}
\figcaption{(a) Calculated postshock temperature compared to the {\it RXTE} and {\it Swift} data (same symbols as in Fig.~1). {\it Dashed curve}: $\eta_{\rm inj}$=10$^{-5}$. {\it Solid and dotted curves}: $\eta_{\rm inj}$=1.4$\times$10$^{-4}$ with and without Alfv\'en wave heating of the shock precursor, respectively. All calculations assume $\alpha_B$=2, $\eta_{\rm mfp}$=20 and $f_{\rm esc}$=0.25. (b) Nonthermal energy fraction ({\it thin curves}) and escape energy fraction $\epsilon_{\rm esc}$ ({\it thick curves}) for the same parameters as in panel (a). For $\eta_{\rm inj}$=10$^{-5}$, $\epsilon_{\rm esc}$ is $<$10$^{-2}$. (c) Subshock and total compression ratios ({\it dashed curves}, left axis) and normalized postshock pressures (right axis) in thermal ($P_{\rm th}$) and nonthermal ($P_{\rm CR}$) particles ($P_{\rm tot}$=$P_{\rm th}$+$P_{\rm CR})$ for $\eta_{\rm inj}$=1.4$\times$10$^{-4}$ and Alfv\'en wave heating. \label{fig3}}
\end{center}
\end{figure}

\end{document}